\documentclass{iaus}
\usepackage{graphicx}

\title[JD 11.~~Radio halos/relics and Cosmic Rays] 
{Searching for radio relics and halos. Their role in the formation and acceleration of extragalactic cosmic rays.}

\author[Gizani A.B. Nectaria]   
{Gizani$^1$ A.B. Nectaria}

\affiliation{$^1$ School of Natural Sciences and Technology, Hellenic Open University \\ 
Tsamadou 13-15 \& Ag. Andreou, 26222 Patra, Greece\\ email: {\tt ngizani@eap.gr} \\[\affilskip]}

\pubyear{2008}
\volume{275}  
\pagerange{119--126}
\jname{IAU Symposium 275:Jets at all scales}
\editors{A.C. Editor, B.D. Editor \& C.E. Editor, eds.}
\begin{document}

\maketitle

\begin{abstract}

We search for extended regions of radio emission not associated with Active Galactic Nuclei, known as `relics', `halos' and 'mini halo's, in a sample of 70 Abell clusters for which we have radio, optical and X-ray data. AGN can produce particle bubbles of non-thermal emission, which can restrict cosmic rays. Hence radio relics and (mini) halos could be forming as a result of the confinement of cosmic rays by these bubbles. 
We are probing the role that intracluster magnetic fields (using Faraday rotation measure and inverse compton arguments), mergers (through radio/X-ray interactions), cooling flows (X-ray data), radio jets/shocks as well as radio (mini) halos/relics play in the formation, acceleration and propagation of cosmic rays. 
For the current study we have selected two powerful nearby radio galaxies from our sample: Hercules A and 3C\,388. We report on the work in progress and future plans. 

\end{abstract}

\firstsection 

\section{Introduction}

Diffuse non-thermal radio emission of relativistic particles not directly associated with cluster galaxies has been classified as haloes, relics and minihalos. These low surface brightness features are thought to be formed via reacceleration of a relic population of relativistic electrons, or proton proton collisions with the intracluster medium (ICM) (see Ferrari et al. 2008 \nocite{ferrari}, and references therein).   

The Auger Collaboration correlates ultra high energy cosmic rays (UHECRs) with radioloud AGN lying near the super galactic plane and their acceleration with turbulent plasma and shocks in their (inner) jets and radio-lobes (e.g. \cite{dremmer}). Radio galaxies are powerful enough to heat and support the cluster gas with injected cosmic-ray protons and magnetic field densities (\cite{ensslin96}); Hardcastle (2010) \nocite{hardcastle}, finds that radiogalaxies should have certain properties in order to accelerate UHECRs and should already be catalogued (i.e. be in the local universe). Jets with particle composition of low energy e$^{-}$/e$^{+}$ and relativistic protons (\cite{leahya}; \cite{leahy}) can also be energetic sources of cosmic rays (CRs). 

Hercules A (z=0.154) is the fourth brightest radiogalaxy (FR1.5) in the sky at low frequencies. We have studied the environment of the AGN in terms of its magnetic field (\cite{gizanix};\cite{gizanir}; see also Gizani (2010a) these proceedings). Our ROSAT X-ray data revealed a cool component of the ICM (contributing to a multiphase intracluster gas) and a dense environment. 

3C\,388 (z = 0.0908) is a relatively small classical double FRII at the centre of a poor cluster with very dense ICM. Our ROSAT HRI observations have revealed warm gas confining the radio lobes (\cite{leahy}). Emission is detected to a radius larger than the radio lobes. 

In the current paper we adopt a cosmology in which H$_{\circ}$ = 65 km s$^{-1}$ Mpc$^{-1}$ and q$_{\circ}$ = 0.

\section{Discussion}

Both radiogalaxies in the optical are cD galaxies. West (1994), suggests that cD galaxies - powerful radiosources are formed through anisotropic mergers (\cite{west}). Neither Hercules A nor 3C\,388 present halos/relics which would be expected as a result of a merger event.

The projected magnetic field closely follows the edge of the lobes, the jets and rings of Her A, suggesting that particle pressure dominates the lobe interior. The magnetic pressure should dominate in the shell region defined by the lobe boundary. Our X-ray analysis showed that the gas thermal pressure is greater than the minimum pressure in the radio lobes by an order of magnitude or so (\cite{gizanix}; \cite{leahy}). This implies that the radio structure of the AGN is confined by the ambient ICM rather than by shocks. The energy supply of the lobes comes mostly from relativistic particles and magnetic fields (little entrainment).  We have also found that magnetic field is below equipartition and therefore unimportant in the lobe dynamics. Hence "invisible" particles (relativistic protons, low energy  e$^{-}$ / e$^{+}$) should dominate (\cite{leahy}). 

 IC arguments suggest $\sim 4.3 \mu$G for Hercules A and $\sim 3.8 \mu$G for 3C\,388. Similar to trapping of CRs in the magnetic field in the Galactic plane they could be trapped in the IC magnetic field. ICM consists of cosmic ray particles, $\mu$G fields and hot gas. Cosmic ray protons with cooling times equal to or larger than the Hubble-time make up a part of ICM. Considering the growth speed of the lobes, synchrotron cooling time is too short for the relativistic electrons to diffuse in them. Both cooling flow clusters have extended radio emission. The short cooling time of the emitting cosmic ray electrons (visible in the spectral index maps, e.g. \cite{gizanir}) and the large extent of the radio sources suggest an ongoing acceleration mechanism in the ICM. Steep spectral indices found in the lobes of Hercules A and 3C\,388 imply short lifetimes of radiating particles and also, to some extent, re-acceleration of the electrons suggesting energy redistribution in the ICM.

The radiative cooling time (in Gyr) in the central regions of the Her A cluster is $\simeq$ 6 Gyrs for the hot phase and 2 Gyrs for the cool phase, while for the  3C\,388 cluster is 3.7 Gyrs. Compared to the age of the universe ($\sim$ 10~Gyrs) the existence of a cooling flow is suggested.

\end{document}